\documentclass[apj]{emulateapj}
\begin{document}
\title{A Fast New Public Code for Computing Photon Orbits in a Kerr Spacetime}
\shorttitle{Analytic Kerr Photon Orbits}
\shortauthors{Dexter and Agol}
\author{Jason Dexter}
\affil{Department of Physics, University of Washington, Seattle, WA 98195-1560, USA}
\email{jdexter@u.washington.edu}
\author{Eric Agol}
\affil{Department of Astronomy, University of Washington, Box 351580, Seattle, WA 98195, USA}
\slugcomment{ApJ, accepted}
\begin{abstract}
Relativistic radiative transfer problems require the calculation of photon trajectories in curved spacetime. We present a novel technique for rapid and accurate calculation of null geodesics in the Kerr metric. The equations of motion from the Hamilton-Jacobi equation are reduced directly to Carlson's elliptic integrals, simplifying algebraic manipulations and allowing all coordinates to be computed semi-analytically for the first time. We discuss the method, its implementation in a freely available FORTRAN code, and its application to toy problems from the literature.
\end{abstract}
\maketitle

\section{Introduction}

Efficient and accurate computation of null geodesics in the vicinity of spinning black holes is important for studies of active galaxies, X-ray binaries, and other accreting black hole systems. The radiated flux from accretion disks mostly originates in the innermost radii, where relativistic effects are important for understanding observations. Proper calculation of the bending of light requires integration along rays \citep{broderick2006}. In general, propagation through the plasma will influence the photon trajectories, leading to non-geodesic paths (\citealt{broderickblandford2003,broderickblandford2004}). However, these effects are mostly important at low frequencies, comparable to the expected plasma and cyclotron frequency. When plasma effects can be neglected, the rays correspond to null geodesics, and these circumstances are assumed throughout this paper.

The first applications of general relativistic radiative transfer to accreting systems were of two main types. \citet{cunningham1975} packaged all radiative effects for optically thick, geometrically thin disks as a transfer function to go from local emissivity to that observed at infinity. \citet{luminet1979} used the simple relationships between impact parameters at infinity and constants of the motion to shoot rays backwards in time from an observer's photographic plate to the object under study. More recently, \citet{viergutz1993} and \citet{beckwithdone2005} considered the so-called emitter-observer problem. That is, given locations of the emitter and the observer, determine the constants of the motion for null geodesics connecting the two. This approach is much more efficient when the source is highly localized, such as an orbiting star or hotspot. Here, backwards ray shooting is impractical since most of the rays miss the target.

Such techniques have been applied to the study of emission lines and spectra from active galactic nuclei (AGN) accretion disks and tori \citep{cadez1998,wu2007} as well as their quasi-periodic oscillations (QPOs) \citep{schnittman2006}. \citet{li2005} used a ray tracing approach to study the spectra of X-ray binaries. \citet{noble2007} created images of galactic center black hole candidate Sagittarius A* (Sgr A*) using axisymmetric general relativistic MHD (GRMHD) simulations, and \citet{bromley2001} studied its polarization from a simplified accretion model. \citet{broderickloeb2006} modeled the frequency dependence of its centroid position, and \citet{reid2008} used ray tracing to compare hot spot accretion models with the observed astrometric motion of its mean position as a function of wavelength. Finally, although the spacetime surrounding neutron stars only asymptotically approaches the Kerr metric, using its null geodesics for ray tracing has still found application in modeling spectra of neutron stars \citep{braje2000}.

Despite all of this work, numerical integration of Kerr null geodesics is computationally expensive in certain applications. \citet{rauchblandford} \defcitealias{rauchblandford}{RB94} (hereafter RB94) described a method for calculating null geodesics in the Kerr metric semi-analytically using the Hamilton-Jacobi formulation of the equations of motion and used it to study the primary caustic. \citet{bozza2008} used a similar method to investigate caustics of all orders, building on earlier analytic work \citep{bozza2002}. \citet{fanton1997} used a fast analytic version for creating line profiles and accretion disk images, and \citet{agolphd} applied this method to the case of polarization from thin disk accretion. \citet{falcke} went on to use this code along with a simple model for the Galactic center black hole to create images of its accretion flow.

All of this work used Legendre's formulation of elliptic integrals (e.g., \citealt{abramstegun}), and treated the $\phi$ and $t$ coordinates numerically, if at all. The tables given in \citet{carlsonQUART,carlson89,carlson91,carlsonDOUBLE} greatly simplify the reductions of the equations of motion to elliptic integrals. The primary aim of this paper is to use Carlson's integrals to calculate {\it all} geodesic coordinates semi-analytically for the first time.



Section \ref{EOM} gives the geodesic equations in Kerr spacetime. Sections \ref{reductions} and \ref{uf} present the reductions to elliptic integrals and the specifics of our implementation. Section \ref{checks} outlines a variety of checks performed to ensure its validity and accuracy, and discusses the speed improvement that should be expected from using an analytic code. Section \ref{implementation} provides an overview of our code for readers who are not interested in all of its detail, and the code is applied to toy problems and test cases in Section \ref{applications}. Finally, Section \ref{discussion} discusses future work both in extending the code and in applying it to more realistic astrophysical situations.

\section{Geodesic Equations of Motion}
\label{EOM}
 
 In Boyer-Lindquist coordinates ($t$,$r$,$\theta$,$\phi$), the Kerr line element can be written,

\begin{eqnarray}
\nonumber ds^2 &=& -\rho^2 \frac{\Delta}{\Sigma^2} dt^2 + \frac{\Sigma^2}{\rho^2}\left ( d\phi -\frac{2ar}{\Sigma^2}dt\right )^2\sin^2\theta+\\ 
 &&\frac{\rho^2}{\Delta}dr^2+\rho^2 d\theta^2,
\end{eqnarray}

\noindent with the definitions

\begin{eqnarray}
\Delta &=& r^2-2r+a^2, \hspace{8pt} \rho^2 = r^2+a^2\cos^2{\theta}, \\
\Sigma^2 &=& (r^2+a^2)^2-a^2\Delta \sin^2{\theta},
\end{eqnarray}

\noindent where $a$ is the angular momentum of the black hole and we use units with $G=c=M=1$.

\citet{carter68} demonstrated the separability of the Hamilton-Jacobi equation for geodesics,

\begin{equation}
  -2 \hspace{2pt}\frac{\partial S}{\partial \lambda} = g^{\mu \nu} \frac{\partial S}{\partial x^\mu}\frac{\partial S}{\partial x^\nu},
\end{equation}

\noindent where $S$ is Hamilton's principal function (the classical action) and $\lambda$ is an affine parameter. The separation reduces the equations of motion to quadratures \citep{chandrasekhar83} relating the coordinates $r$ and $\theta$:

\begin{equation}
\int^r \frac{dr}{\sqrt{R}} = \int^\theta \frac{d\theta}{\sqrt{\Theta}},
\end{equation}

\noindent where 

\begin{eqnarray}
R=[(r^2&+&a^2)E-aL_z]^2-\\\nonumber
  &\Delta&[\mathcal{Q}+(L_z-aE)^2+\delta_1 r^2]\\
\Theta=\mathcal{Q}-&[&a^2(\delta_1-E^2)+L_z^2 \csc^2 \theta ] \cos^2{\theta};
\end{eqnarray}

\noindent and the constants of the motion are the angular momentum about the black hole spin axis, $L_z$, the energy, $E$, and Carter's constant $\mathcal{Q}$. $\delta_1 = 0\hspace{2pt}(1)$ for null (timelike) geodesics.

The equations of motion for the cyclic coordinates are

\begin{eqnarray}
t &=& \lambda E + 2 \int^r r[r^2E-a(L_z - aE)]\frac{dr}{\Delta \sqrt{R}} \\
\nonumber \phi &=& a \int^r[(r^2+a^2)E-aL_z]\frac{dr}{\Delta \sqrt{R}}\hspace{4pt} +\\  &\qquad&\qquad\qquad\qquad\int^\theta (L_z \csc^2 \theta - aE)\frac{d\theta}{\sqrt{\Theta}},
\end{eqnarray}

\noindent with
\begin{equation}\label{leqnr}
\lambda = \int^r \frac{r^2}{\sqrt{R}}\hspace{2pt} dr + a^2 \int^\theta \frac{\cos^2{\theta}}{\sqrt{\Theta}}\hspace{2pt} d\theta.
\end{equation}

The signs of the integrals in $r$ and $\theta$ are independent and arbitrary, but are fixed for a given geodesic. It may seem odd that these equations lend themselves to the choice of $r$ or $\theta$ as independent variable to determine the cyclic coordinates $t$ and $\phi$. However, this is the natural outcome of the separation of the Hamilton-Jacobi equation.

\section{Reduction to Carlson Integrals}
\label{reductions}
In reducing the equations of motion from the previous section, we follow closely the treatment given in Appendix A of RB94. First change variables to ($t$, $u$, $\mu$, $\phi$) with $\mu = \cos{\theta}$, $u = 1/r$. This set is more useful computationally, since the location of an observer at infinity is mapped to $u=0$. The domain of $u$ is then $0 \le u \le u_+ \le 1$, where $u_+$ is the location of the event horizon. Similarly, $-1 \le \mu \le 1$. Then the definitions $q^2 \equiv Q/E^2$, $l \equiv L_z/E$, and $\gamma \equiv E/m$ put the equations of motion in dimensionless form. The integral equation relating $u$ and $\mu$ is

\begin{equation}\label{imuiu}
s_\mu \int \frac{d\mu}{\sqrt{M(\mu)}} = s_u \int \frac{du}{\sqrt{U(u)}};
\end{equation}

\noindent where 

\begin{eqnarray}
M &=& q^2+(\tilde{a}^2-q^2-l^2)\mu^2-\tilde{a}^2\mu^4\\ 
U &=& (1-\gamma^{-2})+2\gamma^{-2}u+(\tilde{a}^2-q^2-l^2)u^2 + \nonumber\\
&\qquad&\qquad\qquad 2[(a-l)^2+q^2]u^3-a^2q^2u^4,
\end{eqnarray}

\noindent and  $\tilde{a}^2=(1-\gamma^{-2})a^2$. This paper only considers null geodesics, so that $\gamma^{-2}=0$ throughout. The arbitrary signs have been written explicitly, and are chosen to be $s_x=\mathrm{sign}(\dot{x})$, where a dot refers to a derivative with respect to affine parameter. This is done so that both sides of (\ref{imuiu}) are always positive. The equations for the other coordinates become

\begin{eqnarray}
\nonumber t-t_0&=&s_\mu\int a^2\mu^2\frac{d\mu}{\sqrt{M}} \hspace{4pt}+ \\
      \label{teqn} &\qquad& s_u\int \frac{2a(a-l)u^3+a^2u^2+1}{u^2(u/u_+-1)(u/u_--1)} \frac{du}{\sqrt{U}} \\
\nonumber \phi-\phi_0&=&s_\mu\int \frac{l\mu^2}{1-\mu^2}\frac{d\mu}{\sqrt{M}}\hspace{4pt} + \\
     \label{peqn} &\qquad& s_u\int \frac{2(a-l)u+l}{(u/u_+-1)(u/u_--1)} \frac{du}{\sqrt{U}},
\end{eqnarray}

\noindent where $u_{\pm} = [1\pm\sqrt{1-a^2}]^{-1}$. The limits of integration have been omitted due to complications in accounting for turning points. This is discussed in more detail below.

Given initial and final values of $u$ and $\mu$, we can compute $t$ and $\phi$. Since the $\mu$ integral is easier to invert and this method is of more general utility, $u$ is taken as the independent variable and the goal is to solve for $\mu_f$ given $\mu_0$, $u_0$ and $u_f$. In certain applications it is more convenient to choose $\mu$ as the independent variable. For example, in the case of thin disk accretion we know the inclination angle as well as the value of $\mu$ where the geodesic intersects the disk. Section \ref{uf} gives solutions for $u_f$ given $\mu_0$, $\mu_f$ and $u_0$ to handle these cases.

\begin{deluxetable*}{lllclr}
\tabletypesize{\scriptsize}
\tablecolumns{6}
\tablewidth{0pt}
\tablecaption{Reduction of $I_u$\label{tableiu}}
\tablehead{\colhead{} & \colhead{Case} & \colhead{Parameter Range} & \colhead{Arguments $(a_i,b_i)$;$(f_j,g_j,h_j)$} & \colhead{$u \hspace{2pt}\epsilon \hspace{2pt} [,]$} & \colhead{\citetalias{rauchblandford}}}
\startdata

$1$&Cubic (3 real) & $a=0,q^2+l^2\ge27$ or \hspace{5pt} $u \le u_2$ & $(-u_1,1)$, $(u_2,-1)$, $(u_3,-1)$ & $[0,u_2]^{\tablenotemark{b}}$ & $1$,$3$,$8$,$10$ \\
 & & $a \ne 0,q^2=0,\vert l \vert \ne \vert a$, & & & \\
$2$&Cubic (3 real) & $a=0,q^2+l^2\ge27$ or \hspace{5pt} $u \ge u_3$ & $(-u_1,1)$, $(-u_2,1)$, $(-u_3,1)$ & $[u_3,u_+)^{\tablenotemark{c}}$ & $2$,$4$,$9$,$11$ \\
 & & $a \ne 0,q^2=0,\vert l \vert \ne \vert a$, & & & \\
$3$& Cubic (1 real) & $a=0,q^2+l^2<27$ or $a \ne 0,q^2=0,l \ne a$ & $(-u_1,1)$;$({2u_1[(a-l)^2+q^2]}^{-1}, f/u_1,1)$ & $[0,u_+)$ & $5$,$7$,$12$ \\
$4$& No roots & $q^2=0,l=a$  & ... & $[0,u_+)$ & $6$ \\
$5$& Quartic (2 real) & $a \ne 0$, $q^2 > 0$ & $(-u_1,1)$,$q_s(u_4,-1)^{\tablenotemark{d}}$;$([-a^2 q^2 u_1 u_4]^{-1}, [u_1^{-1}+u_4^{-1}]f,1)$ & $[0,u_+)$ & $13$,$19$ \\
$6$&Quartic (0 real)  & $a \ne 0$, $q^2 < 0$ & $\left(e^{-1/2},\frac{d}{e(h_2-h_1)},h_1^{\tablenotemark{a}}\right),(e^{-1/2},-g_1,h_1^{-1})$ & $[0,u_+)$ & $14$ \\
$7$&Quartic (4 real) & $a \ne 0$, $q^2 \ne 0$, $u \le u_2$  & $(-u_1,1)$, $(u_2,-1)$, $(u_3,-1)$,$(u_4,-1)$ & $[0,u_2]^{\tablenotemark{b}}$ & $15$,$17$ \\
$8$&Quartic (4 real) & $a \ne 0$, $q^2 \ne 0$, $u \ge u_3$  & $(-u_1,1)$, $(-u_2,1)$, $(-u_3,1)$,$(u_4,-1)$ & $[u_3,u_+)^{\tablenotemark{c}}$ & $16$,$18$ \\
\enddata

 \tablenotetext{a}{$h_1$ is found from solving Eq. (\ref{h1}) and selecting one of the two real roots. $d$, $e$ are defined in Eq. (\ref{de}).}
  \tablenotetext{b}{When $u_2 = u_3$, the domain of $u$ is $[0,u_2)$.}
  \tablenotetext{c}{When $u_2 = u_3$, the domain of $u$ is $(u_3,u_+)$.}
  \tablenotetext{d}{$q_s={sign}(q^2)$.}
\end{deluxetable*}

\subsection{Reduction of $I_u$}

Call the left-hand side (LHS) and right-hand side (RHS) of (\ref{imuiu}) $I_\mu$ and $I_u$ respectively, and start with the reduction of $I_u$:

\begin{equation}
  I_u = s_u \int \frac{du}{\sqrt{U(u)}}.
\end{equation}

\noindent Except in the special case with $a=l, q^2=0$, $U(u)$ is either a quartic or cubic and its roots are denoted $u_i$ with $i=1-3,4$, and ordered increasingly. If real, $u_1 < 0$ and in the quartic case, $u_4 > 1$ or $u_4 < 0$. They are of no physical significance. When all roots are real, the allowed regions for the integrand are $u>u_3$ and $u<u_2$ so that $U$ is positive. Thus the roots are the turning points for null geodesics starting outside $u_2$ and inside $u_3$ respectively in both the cubic and quartic cases. There can be no more than one turning point, since the allowed region is bounded on one side either by infinity or the event horizon. When one or both pairs of roots are complex, there is no turning point in $u$. 

Upon encountering a turning point, the sign of $u$ is reversed, so that the total integral is the sum of the integral from $u_0$ to the turning point and that from $u_f$ to the turning point. The idea is to ensure that the integrals in $u$ and $\mu$ monotonically increase along a geodesic. In a sense this allows the independent variable to take the place of the affine parameter, which cannot be used since it is a function of $u$ and $\mu$.

\citet{carlsonQUART,carlson89} contain formulas for evaluation of integrals of the form

\begin{equation}\label{carlsonform}
  [p] = \int_y^x \prod_{i=1}^{5} (a_i+b_i t)^{p_i/2} dt;
\end{equation}

\noindent with all quantities real, $x > y$, and $a_i+b_i t > 0$ for $y < t < x$. The form of a given integral is described by the vector $[p]$, which contains the powers, $p_i$, of the factored roots. Cases with one or two pairs of complex roots are handled in \citet{carlson91,carlsonDOUBLE}, where they are written in terms of real quantities as
  
\begin{equation}\label{carlsoncplex}
 [p] = \int_y^x (f+g t+h t^2)^{p_2/2} \prod_{i=1,4,5} (a_i+b_i t)^{p_i/2} dt
\end{equation}

\noindent for one pair of complex roots or

\begin{equation}
 [p] = \int_y^x \prod_{i=1}^{2} (f_i+g_i t+h_i t^2)^{p_i/2} (a_5+b_5 t)^{p_5/2} dt
\end{equation}

\noindent for two. In using this form, it is assumed that each power $p_i$ of an irreducible quadratic is written twice in the vector $[p]$. In other words, when one pair of roots is complex, $p_2=p_3$. When all roots are complex, $p_2=p_3$ and $p_1=p_4$.

To ensure that $x > y$ in cases where a turning point may be present, integrals are written in pieces involving the relevant turning point, $u_*$, and the number of turning points along the portion of the geodesic being followed, $N_u$ (either $0$ or $1$):

\begin{equation}\label{iutp}
 I_u = s_u \left (\int_{u_0}^{u_*} \frac{du}{\sqrt{U}} - (-1)^{N_u} \int_{u_f}^{u_*} \frac{du}{\sqrt{U}} \right).
\end{equation}

The Carlson papers reduce all elliptic forms to a set of four fundamental integrals, known as the R-functions \citep{numrecipes}, which replace Legendre's integrals of the first, second and third kind. They are all integrals from $0$ to $\infty$ and hence don't require a limit of integration to be a turning point, greatly simplifying complex root cases where no physical turning point is present. This is one of many advantages of Carlson's approach. As is the case for Legendre's formulation, any elliptic integral can be reduced to a sum of Carlson's R-functions. Where Legendre integrals are used in this paper, they are calculated in terms of the R-functions using the formulas in \citet{numrecipes}. The integrals encountered in this paper are always of the form $p=[-1,-1,-1,-1,p_5]$ for quartic cases and $p=[-1,-1,-1,p_5]$ for cubic cases. Thus the form of coordinate integrals in the following will be specified by $p_5$ alone.
 
To maintain as much generality as possible, all integrals are written as above in terms of their roots. In cubic cases the roots are found from solving the cubic equation, while for quartic cases they are found numerically using the routine \texttt{zroots.f} from \citet{numrecipes}. Finally, instead of writing out the explicit formulas from Carlson's papers and going through the algebra separately in each case, we have written routines for each case. This is much simpler and of more general utility, since numerous integrals must be done to calculate the coordinates of a point along a geodesic.

The integral $I_u$ has $p_5 = 0$ and is given by \citet{carlson89} Eq. (2.12) for real roots for cubic cases. Quartic cases are found in \citet{carlsonQUART} Eq. (2.13) for real roots and \citet{carlsonDOUBLE} Eq. (2.36) for all complex roots. The quartic and cubic cases with a single pair of complex roots are given by \citet{carlson89} Eq. (3.8). The necessary arguments to the Carlson routines are listed by case in Table \ref{tableiu}, along with case definitions, appropriate domains of $u$, and the corresponding cases in Appendix A of \citetalias{rauchblandford}.

As can be seen from Table \ref{tableiu}, writing formulas in terms of the roots of $U$ has the advantage of unifying many disparate cases from previous work. Equal roots cases, which describe orbits approaching the unstable circular photon orbits, cannot strictly speaking be treated identically to other real roots cases as shown in the table. Here, integration to the turning point diverges. The code flags for these cases and integrates them directly from $u_0$ to $u_f$, and the arguments listed in the table are still valid. In practice, however, except for the well known Schwarzschild unstable circular orbits with $q^2 + l^2 = 27$, equal roots cases are almost impossible to trigger. This is because the Carlson routines as written maintain accuracy until $\vert u_2 - u_3 \vert \lesssim 10^{-12}$, which is usually more precise than the determination of the imaginary parts of the roots.

For one pair of complex roots, the arguments $f$, $g$ and $h$ are found by setting $U(u)=q_s(u_4-u)(u-u_1)(f+g u+h u^2)$, where $q_s = \mathrm{sign}(q^2)$, and matching powers of $u$. When all roots are complex, setting $U(u)=(f_1+g_1 u+h_1 u^2)(f_2+g_2 u+h_2 u^2)$ yields five non-linear equations for our six unknown coefficients. The degree of freedom is used to simplify the equations, and a sixth degree polynomial is solved numerically for $h_1$:
 
\begin{equation}\label{h1}
 h_1^6 - \frac{c}{\sqrt{e}} h_1^5 - h_1^4 + \sqrt{e}\left[2 \frac{c}{e}-\left (\frac{d}{e}\right)^2\right]h_1^3-h_1^2-\frac{c}{\sqrt{e}} h_1 + 1 = 0,
\end{equation}
 
\noindent where

\begin{equation}\label{de}
 c = a^2-l^2-q^2,\hspace{5pt} d=2[(a-l)^2+q^2],\hspace{5pt} e=-a^2 q^2.
\end{equation}

\noindent The only pair of real solutions to this equation correspond to the values of $h_1$, $h_2$.

As a full example of one of these reductions, consider case 5 from Table \ref{tableiu} with $u_0 < u_f$ ($s_u=1$). This is the Kerr case with no physical turning points. From \ref{carlsoncplex}, we see that $b_1=1$, $b_4=-q_s$, $a_1=-u_1$, $a_4=q_s u_4$, $x=u_f$, $y=u_0$. The sign $q_s$ is used to keep each factor positive. Matching the powers of $U(u)$ as described above gives $f=-q_s/(u_1 u_4 e)$, $g=(u_4+u_1)/(u_1 u_4) f$, $h=1$. Following \citet{carlson91}, we define

\begin{eqnarray}
 X_i &=& \sqrt{a_i+b_i x}, \hspace{6pt} Y_i = \sqrt{a_i+b_i y}, \\
 \xi &=& \sqrt{f+g x+h x^2}, \hspace{6pt} \eta=\sqrt{f+g y+h y^2}, \\
 c_{ij} &=&\sqrt{2 f b_i b_j-g(a_i b_j+a_j b_i) + 2 h a_i a_j}, \\
 M &=& \frac{(X_1 Y_4 + Y_1 X_4)}{x-y} \sqrt{(\xi+\eta)^2-h(x-y)^2}, \\
 L_{\pm}^2 &=& M^2+c_{14}^2 \pm c_{11} c_{44}.
\end{eqnarray}

Then, 

\begin{equation}
 I_u = \frac{4}{\sqrt{\vert e \vert}} R_F (M^2,L_-^2,L_+^2).
\end{equation}

\noindent $R_F$ is computed using the routine from \citet{numrecipes}. Equations for Carlson elliptic integrals with $p_5 \ne 0$ can similarly be found in the Carlson papers listed above.

\subsection{Inversion of $I_\mu$}

Next, the $I_\mu$ integral needs to be inverted to solve for $\mu_f$. As with $U(u)$, the roots of the biquadratic $M(\mu)$, $M_\pm$, determine the physical turning points in $\mu$. When $M_- > 0$, there are four real roots and the orbit cannot cross the equatorial plane. The physical turning points correspond to the two roots with the same sign as $\mu_0$ and are denoted $\mu_\pm = \mathrm{sign}(\mu_0) \sqrt{M_\pm}$. When $M_- < 0$, the physical turning points are $\mu_\pm = \pm \sqrt{M_+}$ and are symmetric about the equatorial plane. We can calculate the number of times the geodesic has crossed a $\mu$ turning point from the magnitude of the $I_u$ integral. This is done by noting that the maximum value of $\int_{\mu_f}^{\mu_+}$ is $\int_{\mu_-}^{\mu_+}$ and its minimum value is zero. In this derivation the integrand $d\mu/\sqrt{M}$, common to all integrals, is omitted. Then, for $s_\mu=1$,

\begin{equation}\label{tpbad}
 \int_{\mu_0}^{\mu_+} + (N-1) \int_{\mu_-}^{\mu_+} \le I_u \le \int_{\mu_0}^{\mu_+} + N \int_{\mu_-}^{\mu_+},
\end{equation}

\noindent where $N$ is the number of turning points reached in $\mu$, and $\mu_\pm$ are the upper and lower turning points in $\mu$. The integrals are written in these pieces so that they are always positive, as required for use with Carlson's integrals. This condition can be written more concisely as

\begin{equation}\label{tp}
  N = \left \lceil \frac{I_u - \int_{\mu_0}^{\mu_+}}{\int_{\mu_-}^{\mu_+}} \right \rceil,
\end{equation}

\noindent where $\lceil \hspace{2pt} \rceil$ is the ceiling function. If $s_\mu=-1$, then the first turning point reached is $\mu_-$. The condition can then be written

\begin{equation}
 -\left [ \int_{\mu_0}^{\mu_-} + (N-1) \int_{\mu_+}^{\mu_-} \right] \le I_u \le -\left [ \int_{\mu_0}^{\mu_-} + N \int_{\mu_+}^{\mu_-} \right].
\end{equation}

\noindent Using $\int_{\mu_0}^{\mu_-} = \int_{\mu_0}^{\mu_+} - \int_{\mu_-}^{\mu_+}$, we can rewrite this in terms of the same integrals used above:

\begin{equation}
 -\int_{\mu_0}^{\mu_+} + N \int_{\mu_-}^{\mu_+} \le I_u \le -\int_{\mu_0}^{\mu_+} + (N+1) \int_{\mu_+}^{\mu_-}.
\end{equation}

\noindent Finally,

\begin{equation}
 N = \left \lfloor \frac{I_u + \int_{\mu_0}^{\mu_+}}{\int_{\mu_-}^{\mu_+}} \right \rfloor,
\end{equation}

\noindent and $\lfloor \hspace{2pt} \rfloor$ is the floor function. To write out the general solution for $I_u = I_\mu$ for arbitrary number of turning points and $s_\mu$, we include coefficients for the various pieces of the $I_\mu$ integral: 

\begin{equation}\label{mufint}
 I_u = \alpha_1 \int_{\mu_0}^{\mu_+} + \alpha_2 \int_{\mu_-}^{\mu_f} + \alpha_3 \int_{\mu_-}^{\mu_+}.
\end{equation}

The coefficients are functions of $s_\mu$ and $N$ determined by writing down specific cases. For example, $\alpha_1$ reflects whether the integration is positive or negative from $\mu_0$ to $\mu_f$ and is easily seen to be $\alpha_1 = s_\mu$. Similarly, $\alpha_2$ reflects whether the last turning point reached is $\mu_-$ or $\mu_+$. Thus the coefficient is $\alpha_2 = s_\mu (-1)^{N}$. The third coefficient is slightly more complicated and turns out to be

\begin{equation}
 \alpha_3 = 2 \left \lfloor \frac{2N+3-s_\mu}{4} \right \rfloor - 1.
\end{equation}

Armed with the number of turning points and the coefficients, we solve for $\mu_f$ by inverting the second integral on the RHS of (\ref{mufint}):

\begin{equation}
  \int_{\mu_-}^{\mu_f} \frac{d \mu}{\sqrt{M}} = \frac{1}{\alpha_2} \left ( I_u - \alpha_1 \int_{\mu_0}^{\mu_+} - \alpha_3 \int_{\mu_-}^{\mu_+} \right).
\end{equation}

\noindent Calling the RHS $I$ and writing out the square root on the LHS for the general case ($a \ne 0$, $q^2 \ne 0$) gives

\begin{equation}
 I = \frac{1}{\vert a \vert}\int_{\mu_-}^{\mu_f} \frac{d\mu}{\sqrt{(M_+-\mu^2)(\mu^2-M_-)}}.
\end{equation}
 
\citet{carlson2005} contains a table for inverting integrals of the form 

\begin{equation}\label{invertform}
  I = \int_y^x \frac{dt}{\sqrt{(a_1+b_1 t^2)(a_2+b_2t^2)}},
\end{equation}

\noindent where all quantities are real, $x > y$, $0 \le y < x$ and either $y=0$, $x=\infty$ or one limit is a root of the integrand. The latter case applies here. 

\subsubsection{$M_- > 0$}\label{mgt0}

When $M_- > 0$, all requirements are met as written, and

\begin{equation}\label{nd}
 \mu_f = \mu_- nd\hspace{2pt}(J,k), \hspace{8pt} J = \mu_+ \vert a \vert I, \hspace{8pt} k^2 = 1 - \frac{\mu_-^2}{\mu_+^2},
\end{equation}

\noindent where $nd\hspace{2pt}(J,k)=1/dn(J,k)$ and $dn$ is a Jacobi-Elliptic function. The $\mu$ integral terms in $I$ are calculated as

\begin{equation}\label{mulegendre}
\int_{\mu_0}^{\mu_f} \frac{d\mu}{\sqrt{M(\mu)}} = \frac{1}{A} F(x,k) 
\end{equation}

\noindent where $F(x,k)$ is Legendre's integral of the first kind \citep{abramstegun}, $x=\sqrt{\frac{M_+ - \mu_0^2}{M_+ - M_-}}$, $A=\vert a \vert \mu_+$ and $k$ is the same as above. The integral between turning points is the complete elliptic integral $K(k)$.

\subsubsection{$M_- < 0$}

When $M_- < 0$, $y < 0$ in (\ref{invertform}) so that (\ref{nd}) is no longer valid. Since the integrand is an even function of $\mu$, we can write

\begin{equation}
 I = \frac{1}{\vert a \vert}\int_{-\mu_f}^{\mu_+} \frac{d\mu}{\sqrt{(\mu_+^2-\mu^2)(\mu^2-M_-)}},
\end{equation}

\noindent which is in the correct form, except that $-\mu_f$ can be negative. This causes no problems. In this case,

\begin{equation}
 \mu_f = \mu_- cn(J,k),\hspace{8pt} J = \sqrt{\mu_+^2-M_-} \vert a \vert I,\hspace{8pt} k^2 = \frac{\mu_+^2}{\mu_+^2-M_-},
\end{equation}

\noindent and we've used $\mu_-=-\mu_+$ for $M_- < 0$. The $\mu$ terms in $I$ are computed the same as in (\ref{mulegendre}), with $k$ defined in (\ref{nd}), $x=\sqrt{1-\frac{\mu_0^2}{\mu_+^2}}$ and $A=\vert a \vert \sqrt{M_+-M_-}$. The integral between the turning points here is twice the complete elliptic integral $K(k)$.

\subsubsection{$q^2 = 0$}

A special case is encountered when $q^2 = 0$. $M(\mu)$ has a double root at $\mu=0$, causing $I_\mu$ to diverge there, and preventing these orbits from reaching the equatorial plane. Hence, they have at most one physical turning point. In this case $I_\mu$ is elementary, and the solution for $\mu_f$ is

\begin{equation}
 \mu_f = \mu_+\hspace{2pt} \mathrm{sech} \left [\hspace{1pt} \vert a \mu_+ \vert \hspace{1pt}I_u - s_\mu s_1 \hspace{2pt} \mathrm{sech}^{-1} (\mu_0 / \mu_+)\right],
\end{equation}

\noindent where $s_1 = \mathrm{sign}(\mu_0)$.

\subsubsection{$a$ = 0}

Finally, when $a=0$ (the Schwarzschild case) the $\mu_f$ integral is again elementary. The solution for $\mu_f$ is then,

\begin{equation}
\mu_f = \mu_- \cos{ \left [ \frac{1}{\alpha_2} \left ( \sqrt{\frac{d}{2}}I_u - \alpha_1 \cos^{-1} {\left(\frac{\mu_0}{\mu_+}\right)} - \alpha_3 \hspace{2pt} \pi \right)\right]}.
\end{equation}

\subsection{$t$ and $\phi$ coordinate integrals}

Given the solution for $\mu_f$, equations for the coordinates $t$ and $\phi$ can be reduced to elliptic integrals as well. Each coordinate is expressed as a sum of integrals over $u$ and $\mu$. As is done above, the $u$ terms are reduced to Carlson's formulation, and the $\mu$ terms to Legendre's.

The $\mu$ integral term in (\ref{teqn}), which we'll denote $T_\mu$, can be written as a single Legendre integral of the 2nd kind. For example, the $\mu_0$ term in the $M_- < 0$ case is reduced as follows:
\begin{eqnarray}
T_\mu &=& \vert a \vert \int_{\mu_0}^{\mu_+} \frac{\mu^2 d\mu}{\sqrt{(\mu_+^2-\mu^2)(\mu^2-M_-)}} \nonumber\\
      &=& \vert a \vert \mu_+ \int_0^x dt \frac{1-t^2}{\sqrt{(1-t^2)(1-t^2-\frac{M_-}{\mu_+^2})}} \nonumber\\
      &=& A \int_0^x dt \frac{1-t^2-\frac{M_-}{\mu_+^2}}{\sqrt{(1-t^2)(1-t^2-\frac{M_-}{\mu_+^2})}}\nonumber + a^2 M_- I_u \\
      &=& A E(x,k) + a^2 M_- I_u,
\end{eqnarray}

\noindent where $E(x,k)$ is the Legendre integral of the second kind with arguments $x$ and $k$ defined in the previous section. The substitution $t=\sqrt{1-\mu_0/\mu_+}$ is made between lines one and two, and $M_-/\mu_+^2$ is added and subtracted from the numerator between lines two and three. In the $M_- > 0$ case, $T_\mu$ is given by the first term of the above formula, with the arguments $A$, $k$, $x$ for that case given with the solution for $\mu_f$ in Subsection \ref{mgt0}.

The $\mu$ term in the $\phi$ component formula (\ref{peqn}) can be reduced to a Legendre integral of the 3rd kind in analogous fashion. For the $M_- < 0$ case we proceed as follows:

\begin{eqnarray}
 \Phi_{\mu} &=& -l I_u + \frac{l}{\vert a \vert} \int_{\mu_0}^{\mu_+} \frac{1}{1-\mu^2}\frac{d\mu}{\sqrt{({\mu_+}^2-\mu^2)(\mu^2-M_-)}} \nonumber\\
          &=& -l I_u +\nonumber\\ &\qquad&\nonumber\frac{l}{\vert a \vert \mu_+} \int_0^x \frac{1}{1-\mu_+^2+{\mu_+}^2 t^2}\hspace{2pt} \frac{dt}{\sqrt{(1-t^2)((1-\frac{M_-}{\mu_+^2})-t^2)}} \\
          &=& -l I_u + \frac{l}{A(1-M_+)} \Pi(n;x,k)
\end{eqnarray}

\noindent where $\Pi(n;x,k)$ is the Legendre integral of the 3rd kind and $n=\frac{\mu_+^2}{1-\mu_+^2}$. The formula for the $M_- > 0$ case is the same, with $n=\frac{M_+ - M_-}{1-M_+}$ and the other arguments defined in Subsection \ref{mgt0} above. Note that we are using the sign convention for $n$ from \cite{numrecipes}, which is opposite that in \citet{abramstegun}.

$T_u$, the $u$ integral term in (\ref{teqn}), is expanded with partial fractions, and after a little algebra is written

\begin{eqnarray}
 T_u &=& s_u u_r \bigg{[} \left ( 2a(a-l)+\frac{a^2}{u_+}+\frac{1}{u_+^3} \right) \int \nonumber \frac{1}{(u/u_+-1)}\frac{du}{\sqrt{U}}\\&\qquad&-\left ( 2a(a-l)+\frac{a^2}{u_-}+\frac{1}{u_-^3}\right)\int\nonumber \frac{1}{(u/u_--1)}\frac{du}{\sqrt{U}}\\&\qquad&+\left ( \frac{1}{u_-^2}-\frac{1}{u_+^2}\right) \int \frac{du}{u \sqrt{U}} + \frac{1}{u_r} \int \frac{du}{u^2\sqrt{U}}\bigg{]},
\end{eqnarray}

\noindent where $u_r \equiv \frac{u_+ u_-}{u_+-u_-}=-(2\sqrt{1-a^2})^{-1}$ is negative. Three of the terms have $p_5=-2$ and one has $p_5=-4$. When a limit of integration is at infinity ($u=0$), this integral blows up, as it should. In practice, the code picks a non-infinite starting radius large enough that the geodesic trajectories from infinity to the starting radius differ negligibly from their flat space counterparts.

Then,

\begin{eqnarray}
\Phi_u&=&s_u u_r \bigg{[} \left (\frac{l}{u_+}+2(a-l) \right)\int\frac{1}{(u/u_+-1)}\frac{du}{\sqrt{U}}-\nonumber\\
 &\qquad&\qquad  \left ( \frac{l}{u_-}+2(a-l)\right)\int\frac{1}{(u/u_--1)}\frac{du}{\sqrt{U}}\bigg{]},
\end{eqnarray}

\noindent where both integrals are already calculated as part of $T_u$.

Finally, the dimensionless affine parameter can also be calculated along the path from (\ref{leqnr}) without any additional integrals:

\begin{equation}\label{leqn}
 \lambda'=s_u\int\frac{du}{u^2\sqrt{U}} + a^2 s_\mu\int\frac{\mu^2 d\mu}{\sqrt{M}}.
\end{equation}

\noindent The first term is from $T_u$ and the second term is $T_\mu$.

Component integrals are calculated the same way as $I_u$ or $I_\mu$ respectively. That is, $\mu$ component integrals are calculated in pieces using the appropriate coefficients as described above while $u$ component integrals are calculated with reference to the physical turning point, if one exists. These are all the integrals required to compute null geodesics in Kerr spacetime. These equations for the $\phi$, $t$ coordinates are written in Boyer-Lindquist coordinates. For certain applications, Kerr-Schild coordinates are used instead. We note here for completeness the analytic transformations between our Boyer-Lindquist coordinates and these Kerr-Schild coordinates ($\tilde{t}$,$\tilde{u}$,$\tilde{\mu}$,$\tilde{\phi}$) \citep{font1999},

\begin{eqnarray}
  \tilde{t}&=& t+\log{\Delta}-u_r \log{\left(\frac{1-u[1+\sqrt{1-a^2}]}{1-u[1-\sqrt{1-a^2}]}\right)},\\
  \tilde{\phi}&=&\phi-a \hspace{2pt} u_r \log{\left(\frac{1-u[1+\sqrt{1-a^2}]}{1-u[1-\sqrt{1-a^2}]}\right)}, \\
  \tilde{u}&=&u, \\
  \tilde{\mu}&=&\mu.
\end{eqnarray}

\noindent The transformations are valid outside the event horizon, where $\Delta$ and the numerator of the other log terms are positive.

\begin{deluxetable*}{lcccccc}
\tabletypesize{\scriptsize}
\tablecolumns{7}
\tablewidth{0pt}
\tablecaption{Solution for $u_f$\label{tableuf}}
\tablehead{\colhead{} & \colhead{$u_f^{\tablenotemark{a}}$} & \colhead{$J^{\tablenotemark{c}}$} & \colhead{$m_1$} & \colhead{$c_1$} & \colhead{$c_2$} & \colhead{$c_3$}}
\startdata
1&$u_1+u_{21} cd^2 J $  & $c_1[I_\mu-I_u(u_0,u_2)]$ & $\frac{u_{32}}{u_{31}}$ & $\frac{\sqrt{u_{31} d}}{2}$ & ... & ...  \\
2&$u_1+u_{31} dc^2 J $  & $c_1[I_\mu+I_u(u_3,u_0)]$ & $\frac{u_{32}}{u_{31}}$ & $\frac{\sqrt{u_{31} d}}{2}$ & ... & ...  \\
3&$\frac{c_2+u_1-(c_2-u_1) cn J}{1+cnJ}$ & $c_1[I_\mu+I_u(u_1,u_0)]$ & $\frac{1}{2}+\frac{6u_1+c_3}{8c_2}$ & $\sqrt{2dc_2}$ & $\sqrt{u_1(3u_1+c_3)}$ & $\frac{a+l}{a-l}$ \\
4 & ... & ... & ... & ... & ... & ... \\
5 & $\frac{u_4 c_5+q_s u_1 c_4 - (q_s u_4 c_5-u_1 c_4) cn J}{(c_4-q_s c_5) cn J + q_s c_4 + c_5}^{\tablenotemark{b}}$ & $c_1[I_\mu+I_u(u_b,u_0)]$ & $q_s \frac{(c_4+q_s c_5)^2-(u_4-u_1)^2}{4c_4c_5}$ & $\sqrt{e c_4 c_5}$ & ... & ...  \\
6  & $c_3+\frac{n(1+c_2^2) sc J}{1-c_2 sc J}$ & $s_uc_1[I_\mu + I_u(c_3,u_0)]$ & $\left(\frac{c_4-c_5}{c_4+c_5}\right)^2$ & $\frac{\sqrt{e}}{2}(c_4+c_5)$ & $\sqrt{\frac{4n^2-(c_4-c_5)^2}{(c_4+c5)^2-4n^2}}$ & $m+c_2n^{\tablenotemark{d}}$ \\
7 & $\frac{u_2-c_2u_3 sn^2 J}{1-c_2 sn^2 J}$ & $c_1[I_\mu-I_u(u_0,u_2)]$ & $\frac{u_{41}u_{32}}{u_{42}u_{31}}$ & $\frac{\sqrt{e c_3}}{2}$ & $\frac{u_{21}}{u_{31}}$ & $u_{42}u_{31}$ \\
8 & $\frac{u_3-c_2u_2 sn^2 J}{1-c_2 sn^2 J}$ & $c_1[I_\mu+I_u(u_3,u_0)]$ & $\frac{u_{41}u_{32}}{u_{42}u_{31}}$ & $\frac{\sqrt{e c_3}}{2}$ & $\frac{u_{43}}{u_{42}}$ & $u_{42}u_{31}$ \\
\enddata
  \tablenotetext{a}{$sn J = sn(J,m_1)$, $cn J = cn(J,m_1)$, $sc J = sn(J,m_1)/cn(J,m_1)$ and $m_1 = 1-k^2$ is used instead of $k$. $u_{xy} \equiv u_x-u_y$.}
  \tablenotetext{b}{$q_s = \mathrm{sign}(q^2)$. If $q_s = 1$ then $u_a = u_4$, $u_b = u_1$. Otherwise, $u_a=u_1,u_b=u_4$.}
  \tablenotetext{c}{$I_u(y,x) = s_u \int_y^x \frac{du}{\sqrt{U(u)}}$.}
  \tablenotetext{d}{Complex roots are written as $m\pm in$, $p\pm ir$ and are ordered so that $m > p$ and $n > 0$.}
\end{deluxetable*}

\begin{table*}
\caption{\label{tablec}Auxillary Constants used in table \ref{tableuf}}
\begin{scriptsize}
\begin{center}
\begin{tabular}{lcc}
        \tableline
	\tableline
  &  $c_4$ & $c_5$\\
        \tableline
5 & $\sqrt{(m-u_4)^2+n^2}$  & $\sqrt{(m-u_1)^2+n^2}$ \\
6 & $\sqrt{(m-p)^2+(n+r)^2}$ & $\sqrt{(m-p)^2+(n-r)^2}$ \\
	\tableline
\end{tabular}
\end{center}
\end{scriptsize}
\end{table*}

\section{Solution for $u_f$}
\label{uf}
For some applications, it is preferable to use $\mu$ as the independent variable and solve for $u_f$ given $u_0$. In particular, consider geodesics connecting an observer at infinity with a thin, equatorial accretion disk. The initial polar angle is the inclination of the observer. The final polar angle is $\pi/2$ ($\mu_f=0$), and we solve for the radial coordinate where the ray intersects the disk. This method, however, is of less general utility than that described above. Even in simple geometries, the number of turning points in $\mu$ along a geodesic is not known in advance as it must be to use $\mu$ as the independent variable. One way around this is to calculate all geodesics connecting the observer with the disk for a fixed number of $\mu$ turning points \citep{cunnbard1973,viergutz1993}.

The approach in solving for $u_f$ is the same as in solving for $\mu_f$. The integral $I_\mu$ is computed as a Legendre integral of the first kind. Given the number of turning points, $I_\mu$ is computed in pieces as shown above using the coefficients $\alpha_{1,2,3}$.

After finding $I_\mu$, we invert $I_u$. This inversion ranges from relatively straightforward to algebraically formidable. As examples, we discuss cubic and quartic real roots cases in detail. Table \ref{tableuf} gives the solution for $u_f$ in all cases. This problem was first addressed by \citet{agolphd} and the solutions here are from its Table 5.2 with some modification. 

For our first example, consider the first two cases of Table \ref{tableiu} where there are three real roots. The integral to invert is 

\begin{equation}\label{rtp}
  I_\mu = s_u\left (\int_{u_0}^{u_+} \frac{du}{\sqrt{U(u)}} \pm \int_{u_f}^{u_+} \frac{du}{\sqrt{U(u)}}\right ),
\end{equation}

\noindent where $u_+$ is the relevant turning point: $u_2$ or $u_3$. Denote the first term $I_{u_+}$, and write the second term in terms of the roots of the integrand:

\begin{equation}
I_\mu - \hspace{2pt}I_{u_+} = \pm \frac{s_u}{\sqrt{d}}\int_{u_f}^{u_+} \frac{du}{\sqrt{(u-u_1)(u-u_2)(u-u_3)}},
\end{equation}

\noindent where $d=2[(a-l)^2+q^2]$. This can be put in the form (\ref{invertform}) with the substitution $z = \sqrt{u-u_1}$:

\begin{equation}\label{zform}
\pm I = \int_{\sqrt{u_f-u_1}}^{\sqrt{u_+-u_1}} \frac{dz}{\sqrt{[z^2+(u_1-u_2)][z^2+(u_1-u_3)]}},
\end{equation}

\noindent where 

\begin{equation}
I \equiv \frac{\sqrt{d}}{2}\left ( \hspace{2pt}I_\mu - \hspace{2pt}I_{u_+} \right),
\end{equation}

\noindent and $I_{u_+}$ is determined from the same Carlson formulas as for $I_u$ above. 

Comparing (\ref{zform}) with (\ref{invertform}), we see that $a_1 = u_1-u_2$, $a_2 = u_1-u_3$ and $b_1 = b_2 = 1$. If $u_+ = u_3$, then the limits of integration must be switched, since by definition $x > y$. These integrals correspond to the third row, third and fourth columns of Table 1 from \citet{carlson2005}, and the solutions for $u_f$ are

\begin{eqnarray}
  u_f &=& u_1 + (u_2-u_1)\hspace{2pt} cd^2(J,k), \hspace{12pt} u_0 \le u_2 \\
      &=& u_1 + (u_3-u_1)\hspace{2pt} dc^2(J,k), \hspace{12pt} u_0 \ge u_3,
\end{eqnarray}

\noindent with

\begin{eqnarray}
 J &\equiv& \sqrt{u_3-u_1}\hspace{2pt} I, \qquad k^2 = \frac{u_2-u_1}{u_3-u_1}, \\ \nonumber
 dc(J,k) &\equiv& \frac{dn(J,k)}{cn(J,k)},\qquad cd(J,k) \equiv \frac{cn(J,k)}{dn(J,k)},
\end{eqnarray}

\noindent where $cn$, $dn$ are Jacobi-Elliptic functions. Note that the result does not depend on whether or not a turning point has been reached, since both $cn$ and $dn$ are even in $J$.

When $U(u)$ has four real roots and $u \le u_2$, 

\begin{equation}
I_\mu - \hspace{2pt}I_{u_+} = \pm \frac{s_u}{\sqrt{e}}\int_{u_f}^{u_2} \frac{du}{\sqrt{(u-u_1)(u-u_2)(u-u_3)(u_4-u)}},
\end{equation}

\noindent where $e=\vert a q \vert$. With the substitution $z=\sqrt{\frac{u_2-u}{u_3-u}}$, this becomes

\begin{equation}
\pm I = \int_{0}^{\sqrt{\frac{u_2-u_f}{u_3-u_f}}} \frac{dz}{\sqrt{[z^2+(u_1-u_2)][z^2+(u_1-u_3)]}},
\end{equation}

\noindent where 

\begin{equation}
I \equiv \frac{\sqrt{e}}{2} \left(\hspace{2pt}I_\mu - \hspace{2pt}I_{u_+}\right).
\end{equation}

Again comparing with (\ref{invertform}) and using \citet{carlson2005}, we find

\begin{equation}
 u_f=\frac{u_3(u_2-u_1) sn^2 - u_2(u_3-u_1)}{(u_2-u_1)sn^2-(u_3-u_1)},
\end{equation}

\noindent where

\begin{eqnarray}
 sn &=& sn(J,k), \qquad J=\sqrt{(u_4-u_2)(u_3-u_1)}I \\ k^2 &=& \frac{(u_4-u_3)(u_2-u_1)}{(u_4-u_2)(u_3-u_1)}.
\end{eqnarray}

Again the result is independent of whether or not a turning point is present. In (\ref{rtp}) above, the sign of the second term on the RHS depends on whether a turning point is present. This allows us to determine the number of turning points in $u$.

When complex roots are present, the reduction to standard form (\ref{invertform}) is much more difficult. It is discussed in \citet{erdelyi1981}, and relevant formulas for the inversion can be found there and in \citet{byrdfriedman}. In particular, our cases 3,5 are from \citet{byrdfriedman} equations 239.00 (p86) and 259.00, 260.00 (p133,135). Our formula for case 6 is based on \citet{erdelyi1981} Table 2, p310-311. The intricacy of these reductions demonstrates the advantage of Carlson's method. The computation of integrals is equally efficient with complex or real roots. Unfortunately, when inversion is required, \citet{carlson2005} is only a somewhat more compact version of Legendre's original notation and offers no real advantage over previous work.

\section{Code checks and speed tests}

\label{checks}

Using the solution for $\mu_f$, the equality of $I_\mu$ and $I_u$ has been checked to machine accuracy (at least 14 significant digits in all cases). Once found, $\mu_f$ can be used as an input to recover $u_f$. In this way, the two routines have been shown to agree in all cases. Precision in the calculation is limited by error in the determination of the roots of $U(u)$. An advantage of using Carlson's formulation is that all component integrals are computed without reference to the complex roots, which often have less numerical precision than real ones. Formulas for $u_f$, discussed in Section \ref{uf}, are also written in terms of real quantities leading to higher accuracy.

Certain special cases can be integrated analytically for all components, providing independent checks on component integral formulas. These include $\mu$ cases with $q^2=0$, $u$ cases with $q^2 = 0$, $l=a$ and $u$ cases with equal physical real roots, corresponding to unstable circular photon orbits. In all of these cases, the component integral formulas above agree with the analytic results to machine accuracy. The component integral formulas above also reduce to those derived separately for the Schwarzschild case, $a=0$. All of the formulas given here for $\mu_f$ agree with those in tables 1-2 of \citetalias{rauchblandford}. The Schwarzschild formulas were also tested against the approximate formulas given by \citet{belo2002}.

Further, the implementations of Carlson's integral tables have been checked extensively using the Mathematica \texttt{NIntegrate} function. The same is true of the $t$ and $\phi$ formulas, as well as the individual integral components $I_u$, $T_u$, and $I_\mu$, $T_\mu$, $\Phi_\mu$. 

The R-function routines maintain accuracy until $a \lesssim 10^{-5}$ or $q^2 \lesssim 10^{-10}$. If such parameters are encountered, the code will give a warning and set the offending value to zero.

The geodesic computations have been checked against calculations done by the code used in \citet{falcke} and are found to be in excellent agreement. The FORTRAN implementation of our code is found to be faster than that one by a factor of about $5$, due to the fact that our code computes the minimum number of R-functions possible, and shares them between routines when necessary. The code from \citet{agolphd} was found to be $\sim 100$ times faster than numerical integration in the case of tracing geodesics from infinity to a thin disk. This is an optimal problem for an analytic code, since we can solve for the point where $\mu=0$, whereas a numerical code must integrate the geodesic from infinity until it reaches that point, and then zoom in on the intersection to find an approximate solution to the desired accuracy. In addition, this example didn't include the $t$ and $\phi$ coordinates, which are sped up by a much smaller factor than $u$ and $\mu$.

As a lower bound for the speed improvement of our code over numerical integration, a routine was written to integrate the photon four momentum for all coordinates with respect to affine parameter using the implementation of the Bulirsch-Stoer method from \citet{numrecipes}. We then compared the integration of many points along a single geodesic starting from infinity with this numerical code and our analytic one. This is the ideal case for numerical integration, since the intermediate points calculated along the ray are no longer wasted as in the first example. For the case considered with no turning points in $u$ or $\mu$, the analytic code was found to be faster by a factor of $\sim 3$. 

However, our numerical code for integrating geodesics is much simpler than a complete code would have to be. It cannot handle turning points, and requires knowledge of the affine parameter on the ray in order to know where the region of interest in the integration is. In practice, turning points would have to be detected and a scheme for determining the region of interest in affine parameter implemented. Alternatively, a somewhat more complicated scheme such as the Hamiltonian method described in \citet{schnittman2004} could be adopted. In any case, these additions would slow down geodesic computation. We conservatively estimate, then, that the lower bound for the speed advantage of our analytic code over numerical integration is a factor of $\sim 5$. A similar test for the case of integrating down to the thin disk yielded a speed difference of a factor of $\sim 300$, leading to an upper bound on the speed advantage of $\sim 500$, which is in good agreement with the naive estimate of multiplying the speedup found by \citet{agolphd} by the speedup factor between our code and the one presented there. Then the range of speedup that can be expected by using the analytic code described here is a factor between $5-500$ depending mostly on the application, but also on the specific implementation of the numerical integration code.

In addition to being faster, the analytic formulation is much more flexible. It can calculate an arbitrary number of points beginning and ending anywhere on any geodesic, provided that the constants of the motion can be calculated. This is exploited in the thin disk toy models below, where we solve for the point $\mu_f=0$. It could also allow, for example, a calculation of Compton scattering by tracing rays out from every point on a geodesic, and computing the scattered intensity into that point as a separate ray tracing computation. In any event, the flexibility inherent to an analytic method could allow for more sophisticated calculations in the future, which wouldn't be possible with a numerical code. The main disadvantage of using an analytic code is that the affine parameter cannot be used as an independent variable, which may be desirable for adaptive integration techniques in radiative transfer applications, for example.

\begin{figure}
\epsscale{1.2}
\plotone{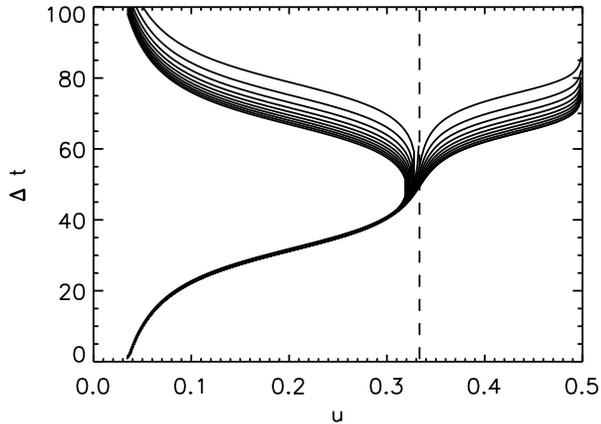}
\caption{\label{prettytime}Change in time vs. radial coordinate in the Schwarzschild metric for geodesics near the circular photon orbit (dashed line), as described in Section \ref{implementation}.}
\end{figure}

\begin{figure}
\epsscale{1.2}
\plotone{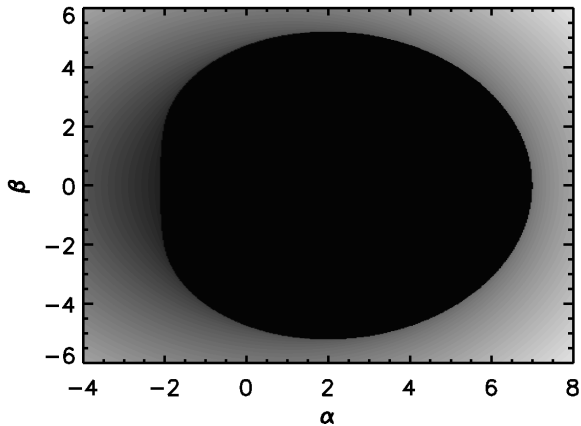}
\caption{\label{bhimage}Image of a near extreme ($a=.998$) Kerr black hole viewed from the equatorial plane. Images intensities are taken to be the affine parameter evaluated upon termination at the black hole or after returning to the starting radius. Intensities are scaled linearly from the minimum value outside the shadow to the maximum.}
\end{figure}

\section{Implementation}
\label{implementation}
This section provides an overview of the various routines used by the code described above, and examples of their use. The README file online covers everything in this section in greater detail. The FORTRAN 77 source file \texttt{geokerr.f} contains the main program as well as the key routines, \texttt{geokerr}, \texttt{geomu}, \texttt{geor} and \texttt{geophitime}, and supporting functions. Inputs are given through command line prompt or a text file. Inputs from previous command line runs may be saved for future use. These inputs include constants of motion for the desired geodesics, initial and final $u$ and initial $\mu$, the number of turning points in $u$ (ignored if the constants do not admit physical turning points), and the sign of $\dot{u}$ and $\dot{\mu}$. Constants of the motion are required and may be specified either as the impact parameters at infinity ($\alpha$, $\beta$), as is most convenient in ray tracing applications, or as the dimensionless angular momentum, $l$, and Carter's constant, $q^2$. When any other information is not provided, the program assumes geodesics which trace out the entire domain of $u$, from the starting point and back or until the event horizon is reached.

The program calls the main subroutine, \texttt{geokerr}, which calls \texttt{geomu} to fill in missing inputs and calculate $\mu_f$. Alternatively, \texttt{geokerr} can solve for $u_f$ using \texttt{geor}. Subsequently, \texttt{geophitime} calculates the $\phi$ and $t$ integrals using the Carlson routines. The program loops over constants of the motion for a chosen initial polar angle and black hole spin. Results are written to standard output by default, and should be redirected from the terminal to a text file in most cases. It is also possible to input the name of the desired output file. The subroutine \texttt{geokerr} encapsulates most of the code functionality, and can fairly easily be adapted to another front end other than the program used here. See the README file accompanying the code or online for more detail.

As an example use of the code, consider tracing rays over a rectangular grid in $-4 \le \alpha \le 8$, $-6 \le \beta \le 6$  for a near extreme black hole, $a=.998$. The observer is at infinity in the equatorial plane ($\mu_0=0$), and $20$ rays will be traced over each dimension. The input file for this situation can be found online.\footnote{http://www.astro.washington.edu/agol/geokerr/exfiles/abgrid.in}

Output is arranged as follows. The constants of the motion are listed for each geodesic in the top line, followed by columns giving $u_f$, $\mu_f$, $\Delta t$, $\Delta \phi$, $\lambda$. The format used for output can be changed with a tiny modification to the source code. See the README file for more details. Plotting the affine parameter evaluated at either the event horizon, or once the geodesic returns to its initial radius, as a function of impact parameters for this data with $160,000$ geodesics produces Fig. \ref{bhimage} as explained below.

For less standard batch runs, it may be necessary to generate the input file from a simple program. Consider a set of geodesics in the Schwarzschild metric ($a=0$) to study the unstable circular photon orbits. The given parameters are chosen to be $u_0=1/30$, $u_f=u_+=.5$, $\mu_0=.9$, $\beta=0$ and an array of values for $\alpha$ near $\sqrt{27}$.

The piece of code to write an appropriate input file is available online.\footnote{http://www.astro.washington.edu/agol/geokerr/exfiles/inputex.f} Plotting the change in time as a function of final radial coordinate produces Fig. \ref{prettytime}.

\begin{figure}
\epsscale{1.0}
\plotone{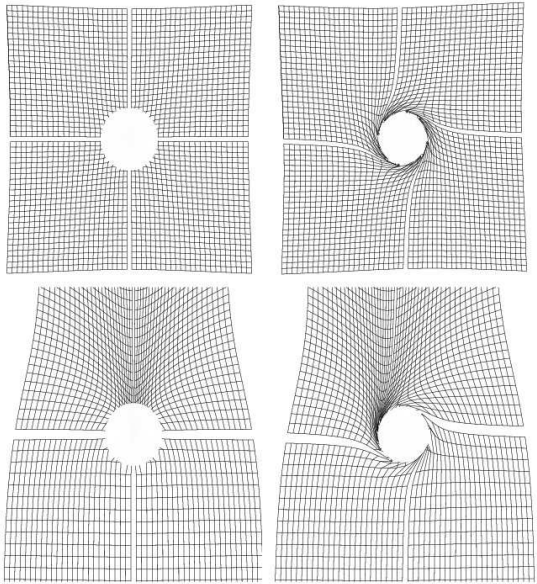}
\caption{\label{grids}Projection of a uniform Cartesian grid in the image plane to the equatorial plane of the black hole for $\mu_0=1$ (top) and $\mu_0=.5$ (bottom). Black hole spin is $a=0$ (left) and $a=.95$ (right), and the area inside the horizon is removed from each image. Compare to Fig. 2 of \citet{schnittmanbertschinger2004}.}
\end{figure}

\section{Applications/Validation}
\label{applications}

We next describe a couple of relatively simple applications of the code to ray tracing problems as further validation and as examples of its utility. The first is the simplest illustration of the black hole shadow, which tests the determination of the roots of $U(u)$ and qualitatively parts of the time integral. Next are examples from the standard model of thin disk accretion. The disk image and simple spectrum from line emission test the routine that solves for $u_f$. The projection of a uniform grid at infinity onto the equatorial plane of the black hole also tests the calculation of $\phi$, and hot spot emission provides a time-dependent test. Finally, spectra and images of synchrotron radiation from spherical accretion quantitatively test our radiative transfer routines.

Ray tracing utilizes the simple relationship between points on an observer's instrument and the constants of motion of null geodesics. Consider the photographic plate at infinity as a function of the impact parameters $\alpha, \beta$, perpendicular and parallel to the black hole spin axis respectively. Images can be created by tracing rays backwards from points on the plate to the black hole. The parameters $\alpha,\beta$ are easily expressed in terms of $q^2,l$ using \citep{cunnbard1973}

\begin{eqnarray}
l &=& -\alpha(1-\mu_0^2)^{1/2} \\
q^2 &=& \beta^2+\mu_0^2(\alpha^2-\tilde{a}^2),
\end{eqnarray}

\noindent so that each point on the observer's photographic plate corresponds to a unique geodesic.

\subsection{Image in Affine Parameter}

As a first application of ray tracing, we can determine the appearance of the simplest possible black hole shadow. The image ``intensities'' are taken to be the affine parameter evaluated at the termination of the geodesic--either when it terminates at the black hole or reaches a turning point and re-emerges to the starting radius. Affine parameter is a good proxy for the emission in this case, since it is related to the proper length along a geodesic, which would be the observed intensity for constant emissivity and neglecting absorption. The dimensionless affine parameter, $\lambda'$, is given by (\ref{leqn}). The equatorial plane result for a Kerr black hole with $a=.998$, to be compared to \citet{bardeen1973} Fig. 6, is shown in Fig. \ref{bhimage}. The image shown here is $400\times400$.

\begin{figure}
\epsscale{1.2}
\plotone{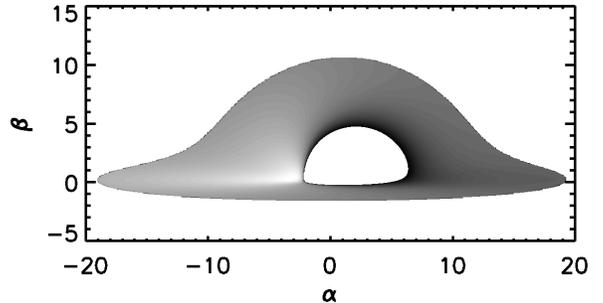}
\caption{\label{diskimage}Image of an optically thick standard relativistic accretion disk around a near extremal black hole (a=.998). The disk has outer radius $r_{out}=18$, and the observer's inclination is $85^{\circ}$.}
\end{figure}

\begin{figure}
\epsscale{1.2}
\plotone{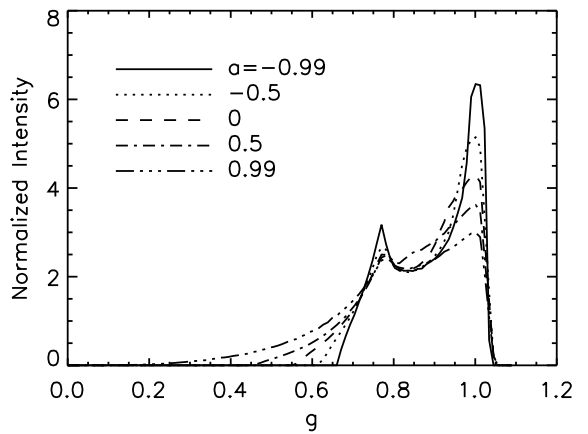}
\caption{\label{lines}Normalized spectra of line emission from a thin accretion disk at an inclination of $30^{\circ}$ for various black hole spins. The emissivity is taken to be proportional to $u_f^2$ between the marginally stable orbit and $R_{out}=15$. Compare to \citet{schnittmanbertschinger2004} Fig. 3.}
\end{figure}

\subsection{Thin Disk Accretion}

The next set of applications imagine the emitting source as an infinitesimally thin disk in the equatorial plane of the black hole (e.g., \citealt{page1974}, \citealt{shaksun1973}). 

\subsubsection{Grid Projection}

The first check of the code for this case is in visualizing the projection of a uniform grid at infinity onto the equatorial plane of the black hole. This is done by solving for the final radius, $u_f$, and azimuth where the geodesic intersects $\mu_f=0$. Then, the new grid points are calculated using pseudo-Cartesian coordinates \citep{schnittmanbertschinger2004}:

\begin{equation}
 x=\sqrt{r^2+a^2} \cos{\phi}, \hspace{5pt} y=\sqrt{r^2+a^2} \sin{\phi}.
\end{equation}

The result of this projection for two different initial observer inclinations and black hole spins is shown in Fig. \ref{grids}, and agrees with Fig. 2 of \citet{schnittmanbertschinger2004}. The gravitational lensing effect can be seen in the pictures with $\mu_0=.5$ as the bunching of grid points behind the black hole, while frame dragging is evident in those with $a=.95$

\subsubsection{Thermal Disk Images}

As a next step, we can use the standard thin disk results for the radial temperature profile (e.g., \citealt{krolikbook}) to produce images of the disk at various inclinations assuming it is optically thick everywhere, so that the intensity is that of a blackbody. Finding the radii of emission from a grid in impact parameters and calculating the intensity at each of these points produces an image of the disk as seen by a distant observer. The result for an inclination of $85^{\circ}$ and black hole spin $a=.998$ is shown in Fig. \ref{diskimage}. The image shows the effects of relativistic beaming of the emission from gas moving towards the observer versus the redshift of that moving away, as well as the bending of the light from gas behind the black hole.

\subsubsection{Line Emission}

Next, following \citet{schnittmanbertschinger2004} and \citet{bromley1997} we consider monochromatic emission from the disk, and give it an inner (outer) radius, $R_{in}=R_{ms}$ ($R_{out}=15$), where $R_{ms}$ is the location of the marginally stable circular orbit (e.g., \citealt{page1974}). The emissivity is weighted by $u_f^2$, physically motivated by the fact that we expect the temperature of gas in the disk to increase with decreasing radius. The observed intensity is computed by exploiting the invariance of $I_\nu / \nu^3$ \citep{mtw},

\begin{equation}
 I_{\nu_0} = g^3 I_{\nu}
\end{equation}

\noindent where $g \equiv \nu_0/\nu$ is the redshift, and $\nu_0$ ($\nu$) is the observed (emitted) frequency. To see the effect of black hole spin on the emission in this case, we calculate $I_{\nu_0}$ as a function of $g$ for several values of $a$ by calculating the intensity of rays at a location with redshift in a certain range of $g$, and integrating them over the photographic plate. The result is plotted in Fig. \ref{lines}, and is in excellent agreement with Fig. 3 of \citet{schnittmanbertschinger2004}. At higher black hole spin, the marginally stable orbit is much closer to the black hole where the redshift is much stronger, leading to a higher relative magnitude and broadening of the low frequency peak (``red wing'').

\begin{figure}
\epsscale{1.2}
\plotone{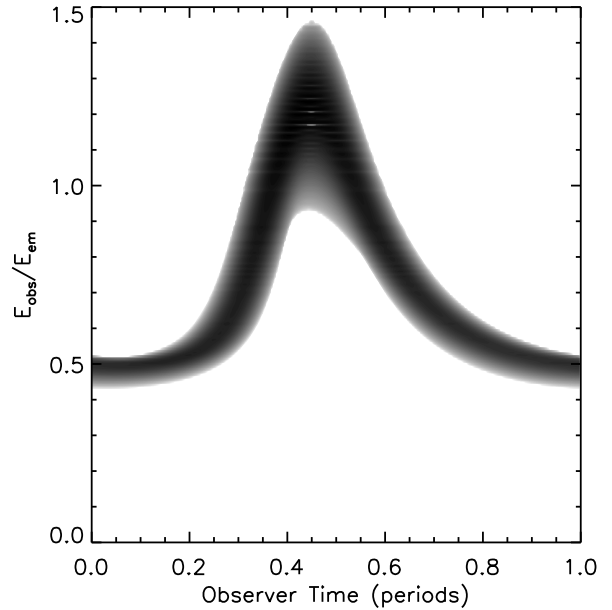}
\caption{\label{spectrogram}Spectrogram of a circular hot spot of radius $R_{spot} = .5$ at the marginally stable orbit of a Schwarzschild black hole. The observer is inclined at $\theta_0=60^{\circ}$. Compare to Fig. 4 of \citet{schnittmanbertschinger2004}.}
\end{figure}

\begin{figure}
\epsscale{1.2}
\plotone{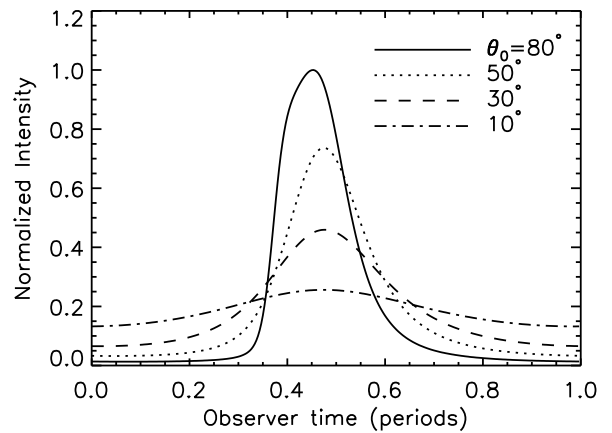}
\caption{\label{lcurve}Light curves of the hot spot described in Fig. \ref{spectrogram} for various inclination angles. Intensities are normalized individually to the integrated intensity over each orbit and scaled to the maximum intensity from all inclinations. Compare to Fig. 6 of \citet{schnittmanbertschinger2004}.}
\end{figure}

\subsubsection{Rotating Hot Spot}

Finally, to test the time-dependence of the code, consider a circular hot spot of finite radius $R_{spot}=.5$ orbiting in the equatorial plane of a Schwarzschild black hole at its marginally stable radius ($R_{ms}=6$). The emissivity of the spot is taken to be Gaussian in the locally flat space near the hot spot (for details, see \citealt{schnittmanphd}),

\begin{equation}
 j(\mathbf{x}) \propto \exp{\left [-\frac{\vert \mathbf{x}-\mathbf{x}_{spot}(t) \vert^2}{2 R_{spot}^2} \right]},
\end{equation}

\noindent where $j$ is the monochromatic emissivity. For some observer coordinate time, $t$, the time delay and azimuthal position from the observer to points on the disk are used to determine where on the photographic plate the separation between geodesic and hotspot are less than $4 R_{spot}$. For these points, the Gaussian emissivity and observed frequency (redshift) are tabulated. Repeating this procedure over a period of the motion gives a time-dependent spectrum, which is shown in Fig. \ref{spectrogram} for an observer inclination of $60^{\circ}$ ($\mu_0=.5$). This figure is in good agreement with Fig. 4 of \citet{schnittmanbertschinger2004}.

Integrating over frequency (redshift), or equivalently over the impact parameters, gives the light curve. Fig. \ref{lcurve} shows the light curves of the hotspot for several inclination angles. As the observer approaches edge-on viewing, the light curve becomes sharply peaked by a combination of the Doppler beaming of the spot as it moves toward the observer and the large gravitational lensing of the spot as it goes behind the black hole. The plot here is in excellent agreement with \citet{schnittmanbertschinger2004}.

\subsection{Radiative Transfer}

In more realistic astrophysical applications, the source is not a delta function at a given inclination, and the intensity along a ray can be written more generally as

\begin{equation}\label{dintense}
I_{\nu_0} = \int_{ray} \left ( \frac{\nu_0}{\nu} \right )^3 dI_\nu.
\end{equation}

If absorption can be neglected, $dI_\nu = j_\nu dl$ where $dl =  -p_\alpha u^\alpha d\lambda$ is the proper length differential measured along the ray, $p^\alpha$ is the photon four-momentum, $u^\alpha$ is the four-velocity of the emitting particle and $\lambda$ is an affine parameter.

\begin{figure}
\epsscale{1.2}
\plotone{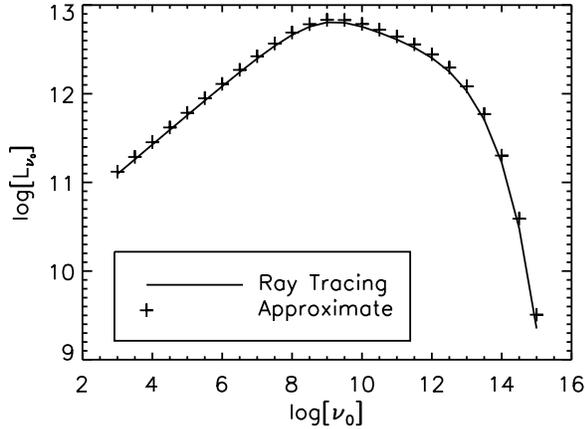}
\caption{\label{tracecompare}The spectrum of synchrotron radiation from optically thin spherical accretion onto a stellar mass black hole. The solid line is the ray tracing result, and the plotted points are the analytic results. The two curves agree to within $5\%$ at low frequencies, where the radiation originates at larger radii and the bending of light should be unimportant.}
\end{figure}

The observed intensity is then

\begin{equation}
 I_{\nu_0} = \int_{\lambda_0}^{\lambda} j_\nu g^2 d \lambda
\end{equation}

\noindent where $j_\nu$ is the emission coefficient in the rest frame of the gas, and $\lambda$ is now the dimensionless affine parameter used above. It is calculated from (\ref{leqn}) and used as the independent variable along the ray. When absorption is included, the solution to the radiative transfer equation between affine parameters $\lambda_0$ and $\lambda$ reads \citep{fuerstwu2004}

\begin{equation}\label{formalradtrans}
  I_{\nu_0}(\lambda) = g^3 I_\nu(\lambda_0) e^{-\tau_\nu(\lambda_0)} + \int_{\lambda_0}^{\lambda} e^{-(\tau_\nu(\lambda ') - \tau_\nu(\lambda_0))} g^2 j_{\nu} \hspace{2pt} d\lambda ',
\end{equation}

\noindent where $\tau_\nu \equiv \int \alpha_\nu dl$ is the optical depth. Throughout this paper we neglect scattering contributions to the emission and absorption coefficients.

\begin{figure}
\epsscale{1.2}
\plotone{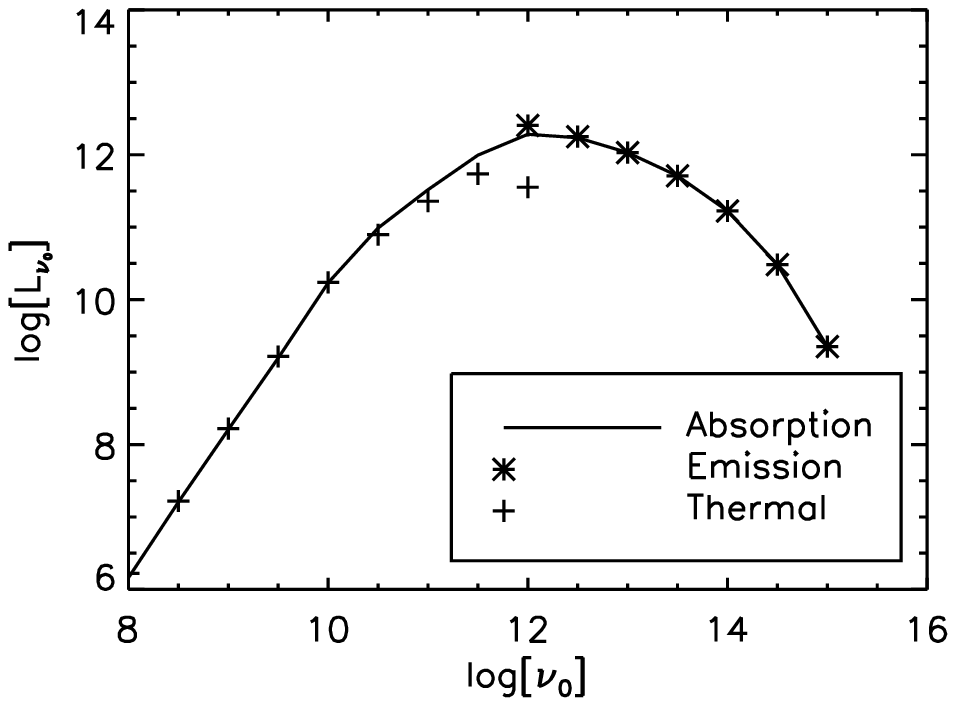}
\caption{\label{abscompare}Spectrum of synchrotron radiation from spherical accretion onto a stellar mass black hole. The solid line is the ray tracing result including absorption. The spectrum is heavily attenuated at $\nu_0 \lesssim 10^{11} \hspace{2pt} \mathrm{Hz}$, and in this region follows the optically thick approximation of thermal emission from the $\tau = 1$ surface. The spectrum agrees well with the emission only model for $\nu_0 \gtrsim 10^{12} \hspace{2pt} \mathrm{Hz}$}
\end{figure}

\subsection{Synchrotron Radiation from Spherical Accretion}

\begin{figure}[ht!]
\begin{center}
\epsscale{1.0}
\plotone{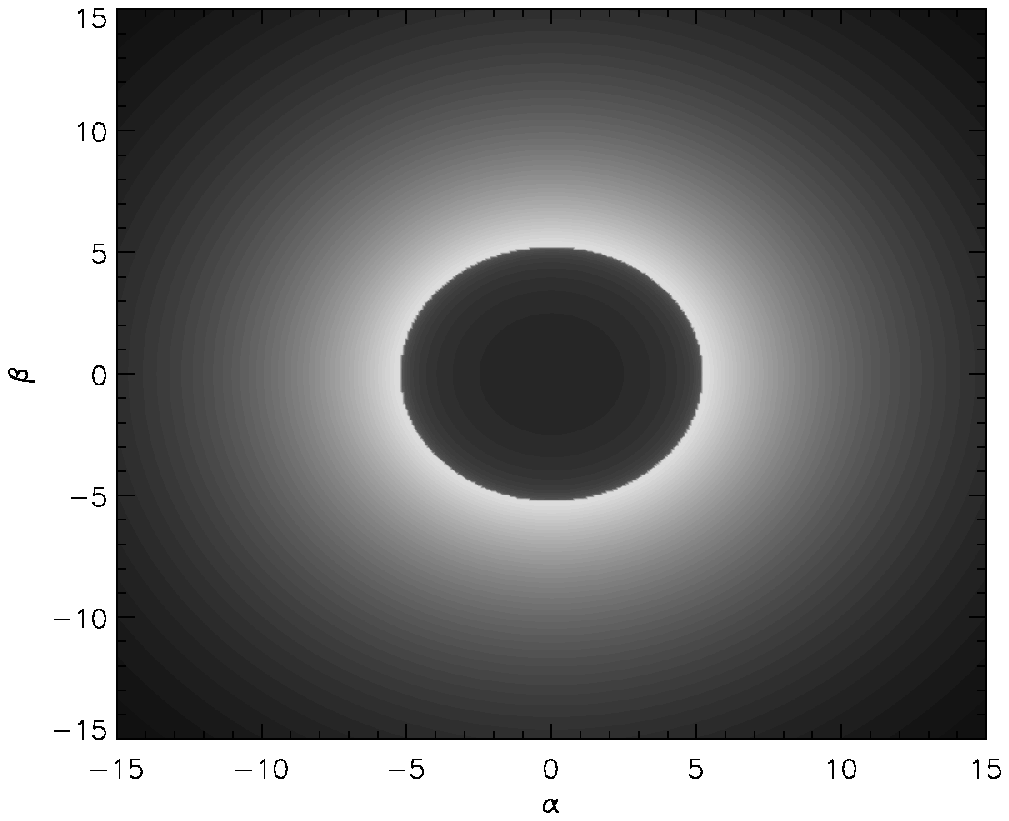}
\epsscale{1.0}
\plotone{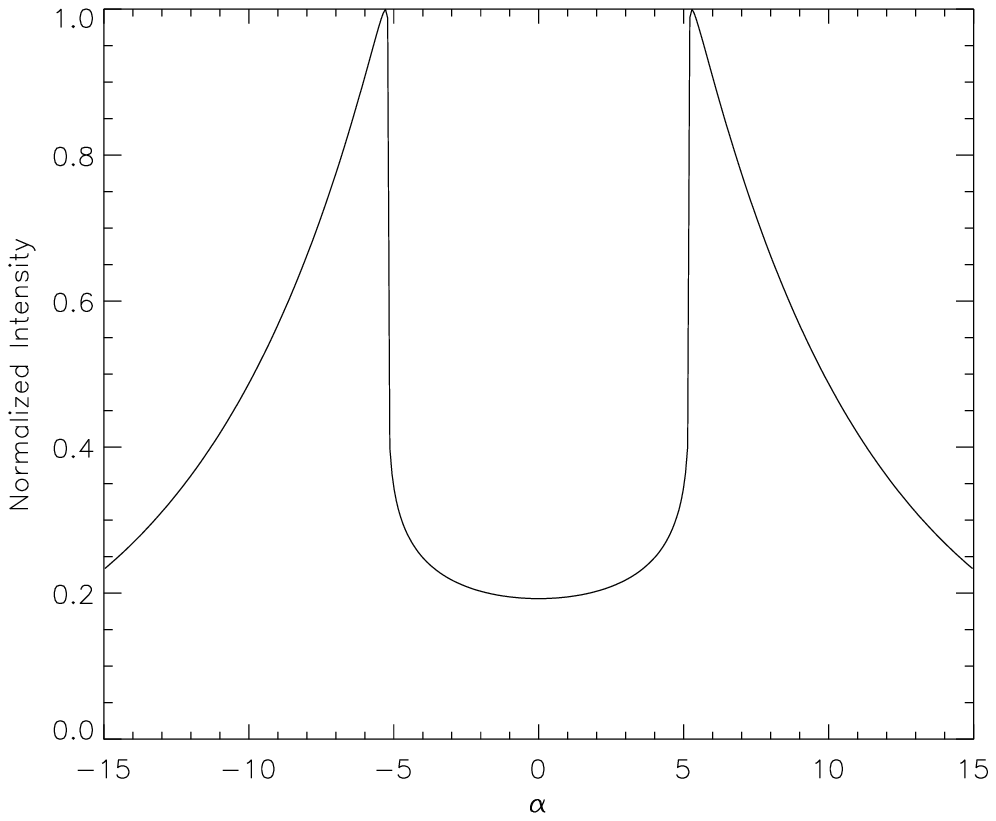}
\caption{\label{imgs} Image of a spherically accreting Schwarzschild black hole at $\nu = 10^{12} \hspace{2pt} \mathrm{Hz}$ as a contour plot and a 1-d profile.}
\end{center}
\end{figure}

The code described above in conjunction with a routine to perform radiative transfer along rays is now applied to the particularly simple case of a stellar mass black hole at rest with respect to the interstellar medium with a temperature at infinity of $10^4$ K and a density at infinity of $1 \mathrm{cm}^{-3}$. Ionized hydrogen accretes onto the black hole, and the magnetic field threading the gas effectively creates collisions, so that the accreting gas can be considered a perfect fluid. In the model, magnetic turbulence establishes an equipartition of magnetic and gravitational energy \citep{zelnov}. Then 

\begin{equation}
  \frac{B^2}{8\pi} = \frac{GM\rho}{r},
\end{equation}

\noindent and cgs units are most convenient in the analytic calculation. We assume an adiabatic equation of state with a piecewise adiabatic index \citep{shapbook}, 

\begin{eqnarray}
\gamma &=& \frac{5}{3},\hspace{20pt} \frac{3}{2} \frac{m_p}{m_e}T \leq 1\nonumber \\
       &=& \frac{13}{9},\hspace{15pt} \frac{3}{2} \frac{m_p}{m_e}T > 1,
\end{eqnarray}

\noindent where $m_p$, $m_e$ are the proton and electron mass and $T$ is the temperature in units of proton rest energy. Then the fluid equations are non-linear, and can be solved numerically \citep{michel} to find the temperature and fluid velocity as functions of coordinate radius.

The dominant form of radiation produced is synchrotron radiation from the inner part of the accreting sphere, where the electrons are ultrarelativistic \citep{shap2}. In this case, the emissivity can be well approximated analytically. \citet{shap1} performed the relativistic radiative transfer by approximating the photons as traveling on null geodesics in Minkowski spacetime, and calculating gravitational redshifts as well as the photon Doppler shifts along these paths. 

Shapiro's formula for the radiated spectrum is

\begin{eqnarray}\label{lint}
 L_{\nu_0} &=& 8 \pi^2 \int_{2m}^{r_*} dr\hspace{2pt} r^2 \times \nonumber\\ &\qquad&\int_{-1}^{\cos{\Theta_c}} d(\cos{\Theta^\prime}) \hspace{2pt} j_\nu \hspace{2pt} \frac{1-v^2}{(1-v\cos{\Theta^\prime})^2} \\
  \nu_0 &=& \nu \hspace{2pt} \frac{\sqrt{(1-v^2)(1-2m/r)}}{1-v\cos{\Theta^\prime}} \nonumber,
\end{eqnarray}

\noindent where $v(r)$ is the proper velocity seen by a stationary observer and

\begin{equation}
 \vert \cos{\Theta_c} \vert = \left [ \frac{27}{4} \left (\frac{2m}{r} \right)^2 \left (\frac{2m}{r}-1 \right)+1\right]^{1/2}
\end{equation}

\noindent is the critical angle at which the light is recaptured by the black hole.

The synchrotron emissivity for thermal, ultrarelativistic electrons averaged over polarization and solid angle assuming isotropic emission in the rest frame is given by \citep{pach},

\begin{eqnarray}
j_\nu(T) &=& \nu \frac{ne^2}{2\sqrt{3}c} \left ( \frac{m_e c^2}{kT}\right )^2 I\left ( \frac{x_M}{\sin{\theta}}\right ), \\
I(x) &\equiv& \frac{1}{4\pi} \int d\Omega \hspace{2pt}\frac{1}{x}\int_0^\infty dz\hspace{2pt} z^2 \exp(-z) F\left(\frac{x}{z^2}\right),
\end{eqnarray}

\noindent with $x_M = \frac{\nu}{\nu_c}$,

\begin{equation}
\nu_c = \left(\frac{3eB}{4 \pi m_e c}\right)\left(\frac{k T}{m_e c^2}\right)^2,
\end{equation}

\noindent and where

\begin{equation}
F(x) \equiv x \int_x^\infty K_{5/3}(y) dy
\end{equation}

\noindent is the synchrotron function. \citet{maha} have approximated $I$($x$) above analytically by matching the asymptotic forms for large and small $x$. They find

\begin{equation}
I \left ( \frac{x_M}{\sin{\theta}} \right) \simeq \frac{4.0505}{x_M^{1/6}} \left(1+\frac{0.40}{x_M^{1/4}}+\frac{0.5316}{x_M^{1/2}}\right)\exp(-1.8899x_M^{1/3}).
\end{equation}

\noindent Note that this function is denoted $I^\prime(x)$ by \citet{maha}, and has a maximum error of $\approx 2.7\%$. The spectrum is calculated by integrating (\ref{lint}) numerically.

To compare with these results, the ray tracing code is used to create an image of the synchrotron radiation from the infalling gas in the same way as done previously with affine parameter. To create an image, one specifies a grid of points in $\alpha$, $\beta$ and calculates $q^2$ and $l$. This fully specifies the geodesic, and we can calculate the spacetime coordinates at which it intersects the accreting gas. The intensity along each geodesic represents a point in the image, which is why it is so important to be able to calculate geodesics rapidly.

The redshift is calculated using \citet{viergutz1993}. Here, the flow is spherically symmetric and, 

\begin{equation}
  g = \left (\gamma e^{-\eta} [1 - e^{\mu_1+\eta} v^r \rho^{-2} r_{sgn} \sqrt{R}]\right)^{-1},
\end{equation}

\noindent with

\begin{equation}
 e^{2\eta} \equiv \Delta \rho^2 \Sigma^{-1}, \hspace{8pt} e^{2 \mu_1} = \rho^2 \Delta^{-1}, \hspace{8pt} \gamma=(1-{v^r}^2)^{-1/2},
\end{equation}

\noindent and $v^r$ is the radial component of the four-velocity. Written in terms of $u$ in the Schwarzschild metric, this simplifies to 

\begin{equation}
  g = \frac{\sqrt{1-2u}}{\gamma[1+s_u(-1)^{N_u} v^u \sqrt{U}]}.
\end{equation}

\noindent When the $u$-component of the four-velocity, $v^u$, vanishes, this reproduces the standard gravitational redshift \citep{hartlebook}. 

We first ignore absorption and compare radiated spectra with the analytic calculation. The result is in Fig.\hspace{2pt}\ref{tracecompare}. \citet{shap2} points out that the synchrotron radiation is dominated by a thin spherical shell of gas with $\nu \simeq \nu_c$. Then the first part of the spectrum, where $L_{\nu_0} \sim \nu^{1/3}$, originates from the outer part of the sphere. The bending of light should be negligible in that region and the ray tracing should agree with the analytic result, which it does to within $\simeq 5\%$. At higher frequencies, the radiation is originating in the innermost radii, and the bending of light becomes significant. The difference is $\simeq 15\%$ at high frequencies.

Next, absorption is included. Fig.\hspace{2pt}\ref{abscompare} compares the spectra calculated with and without absorption. The radiation is heavily attenuated at frequencies $\lesssim 10^{11} \hspace{2pt} \mathrm{Hz}$. At these frequencies, the luminosity is dominated by the innermost optically thin radius, which we take to be the radius where $\tau=1$. Blackbody emission at the temperature of gas at this radius, converted to a luminosity by integrating over impact parameter, is labeled `Thermal' in Fig.\hspace{2pt}\ref{abscompare} and is a decent approximation to the full spectrum when the fluid is optically thick.
 
From $\nu_0 \simeq 10^8 \hspace{2pt} \mathrm{Hz}$ to $\nu_0 \simeq 10^{10} \hspace{2pt} \mathrm{Hz}$, the gas is optically thick everywhere. Then only thermal emission from the outermost radius is seen, and the spectrum follows a Rayleigh-Jeans curve with $L_{\nu_0} \sim \nu_0^2$. From $\nu_0 \simeq 10^{10} \hspace{2pt} \mathrm{Hz}$ to $\nu_0 \simeq 10^{12}$, the innermost optically thin radius is changing, and the luminosity begins to turn over. Starting at $\nu_0 \simeq 10^{12} \hspace{2pt} \mathrm{Hz}$, the gas is optically thin to the synchrotron radiation, and the spectrum reduces to that of emission only (labeled `Emission' in Fig.\hspace{2pt}\ref{abscompare}). This result agrees reasonably well with the assertion made by \citet{shap2} that absorption is negligible when $\nu \gtrsim 10^{11} \hspace{2pt} \mathrm{Hz}$.

Also of interest is the black hole shadow produced by various accretion models \citep{falcke}. Fig.\hspace{2pt}\ref{imgs} shows the shadow of the spherically accreting Schwarzschild black hole as a 2-d contour plot and a 1-d profile. The shadow is produced at $\alpha^2 + \beta^2 = 27$, and is caused by the difference in proper length of geodesics which intersect the horizon and return to infinity, as well as the  blueshift of radiation from infalling gas behind the black hole relative to the red shift of that nearest the observer. The asymmetry in Fig. \ref{bhimage} is not seen here due to the spherical symmetry of the Schwarzschild metric.

\section{Future Work}

\label{discussion}

The code presented here is the first to calculate all coordinates of Kerr null geodesics semi-analytically. This work's natural extension is to timelike geodesics, which involves many more cases, but only straightforward generalizations of the formulas given here \citepalias[Appendix A]{rauchblandford}. The main challenge is that for bound orbits it is difficult to specify the number of radial turning points in advance. Ideally the affine parameter could be used as an independent variable to indicate how far along the geodesic to trace. However, it is a function of $u$ and $\mu$ which cannot be inverted.

\subsection{Advantages of Analyticity}

The main advantages of using a semi-analytic code such as that presented here for tracing geodesics are speed, accuracy and flexibility. The speed increase from our code depends greatly on the application considered. For ray tracing applications, a lower bound is a factor of 5 in the case where all coordinates are being calculated, and the entire ray is being traced. The maximum speed increase is probably a factor between 100-500 in the case where the code is solving for geodesic coordinates at a specific point. 

The importance of speed in tracing geodesics depends on the computational expense of their construction relative to that of the rest of the desired calculation. For the simple radiative transfer applications considered here, time spent computing geodesics dominates in creating Figs. 2,3,4. The construction of geodesics and radiative transfer parts are about equally expensive in creating Figs. 10 and geodesic speed is relatively unimportant in the calculations leading to Figs. 5-9. In the latter cases, this is because the same geodesics can be re-used at many time steps, frequencies or both.

The trend from these toy problems is that the simpler cases benefit most from rapid geodesic calculation. However, there is reason to expect that for more realistic calculations rapid geodesic construction will again be important. Most accretion flows transition from optically thin to thick. To accurately compute radiative transfer from such flows, it is often necessary to take small steps in the vicinity where the optical depth is about unity. This requires calculating extra geodesic trajectories in this region. Since the region where the optical depth changes rapidly depends on frequency, and for a time-dependent accretion model on time as well, new geodesics must be computed at each time step and frequency, and hence they cannot be re-used as is the case for the time-independent, optically thin models considered in almost all our examples.

The precision of our code is also extremely high over a broad range of geodesic parameters. This is currently less important in radiative transfer applications where the dynamical models are uncertain, but it is important in caustic calculations such as those in RB94 and \citet{bozza2008}. Finally, our code can compute arbitrary sections of geodesics in any direction. This flexibility allows extra points to be calculated in regions where the optical depth is changing rapidly or to check convergence on the fly. It may also be useful in a future method for computing Compton scattering, in which rays are traced outwards from each point on the geodesic to calculate the scattered intensity into that point.

Unlike previous analytic work, our code makes no assumption of time-independence or axisymmetry in the accretion flow and is therefore well suited to the geometries used in 3D GRMHD simulations. Computationally expensive observables such as polarization and variability will be much more tractable given the speed and flexibility of this code.

\begin{acknowledgements}
J.D. thanks Jeremy Schnittman for help with the geodesic hot spot model, and the referee, Avery Broderick, whose comments improved this paper and led to substantial rewriting of the code. This work was partially supported by NASA grants 05-ATP05-96 and NNX08AX59H, and a graduate fellowship from the Kavli Institute of Theoretical Physics at the University of California, Santa Barbara under NSF grant PHY05-51164.
\end{acknowledgements}

\end{document}